\documentclass[twoside,12pt]{article}
\usepackage{epsfig}
\usepackage[]{graphicx}

\def\Journal#1#2#3#4{{#1} {#2} (#4) #3 }

\def\NPB{{\em Nucl. Phys.} B}

\def\PRL{\em Phys. Rev. Lett.}

\def\PRD{{\em Phys. Rev.} D}
\def\PRC{{\em Phys. Rev.} C}

\def\ZPC{{\em Z. Phys.} C}
\def\ZPA{{\em Z. Phys.} A}

\def\Fort{\em Fortsch. Phys.}
\def\JHEP{\em JHEP}

\def\IJMPA{{\em Int. J. Mod. Phys.} A}
\def\EPJC{{\em Eur. Phys. J.} C}

\newcommand{\be}{\begin{equation}}
\newcommand{\ee}{\end{equation}}
\newcommand{\bea}{\begin{eqnarray}}
\newcommand{\eea}{\end{eqnarray}}
\newcommand{\nn}{\nonumber}
\def\mev{~{\rm MeV}}
\def\gev{~{\rm GeV}}

\newcommand{\gsim}{\raisebox{-4pt}{$\,\stackrel{\textstyle
                                                         >}{\sim}\,$}}
\newcommand{\req}[1]{(\ref{#1})}
\def\={\,=\,}
\newcommand{\ci}[1]{\cite{#1}}

\begin{document}

\title{ \vspace{1cm} Gereralized Parton Distributions:\\[0.2em]
        Analysis and Applications}
\author{P.\ Kroll\\ 
Fachbereich Physik, Universit\"at Wuppertal, Germany}
\maketitle
\begin{abstract} 
Results from a recent analysis of the zero-skewness generalized parton
distributions (GPDs) for valence quarks are discussed. The analysis bases on a
physically motivated parameterization of the GPDs with a few free
parameters adjusted to the available nucleon form factor data. Various
moments of the GPDs as well as their Fourier transfroms, the quark 
densities in the impact parameter plane, are also presented. The $1/x$ 
moments of the zero-skewness GPDs are form factors specific to Compton
scattering off protons within the handbag approach. The results of the GPD
analysis enables one to predict Compton scattering.\\
Talk presented at the workshop on Lepton Scattering and the Structure
of Hadrons and Nuclei, Erice Italy) September 2004     
\end{abstract}
\section{Factorization and general parton distributions}
In recent years we learned how to deal with hard exclusive reactions
within  QCD. In analogy to hard inclusive reactions like deep
inelastic lepton-nucleon scattering (DIS), the process amplitudes
factorize in partonic subprocess amplitudes, calculable in
perturbation theory, and in GPDs
which parameterize soft hadronic matrix elements. In some cases 
rigorous proofs of factorization exist. For other processes
factorization is shown to hold in certain limits or under certain
assumptions or is just a hypothesis.

The GPDs which are defined by Fourier transforms of bilocal
proton matrix elements of quark field operators \ci{mue1994}, describe
the emission and reabsorption of partons by the proton carrying
different momentum fractions $x_1$ and $x_2$, respectively. Usually
the GPDs are parameterized in terms of the momentum transfer
from the initial to the final proton, $t$, the average momentum fraction
$x$ and the skewness, $\xi$. The latter variables are related to the
individual momentum fractions $x_1$ and $x_2$ by
\be
x_1\= \frac{x+\xi}{1+\xi}\,, \qquad   x_2\= \frac{x-\xi}{1-\xi}\,.
\ee
Although the GPDs are not calculable with a sufficient degree of
accuracy at present, we know many of their properties. Thus, for
instance, they satisfy the reduction formulas
\be
H^q(x,\xi=0,t=0)\= q(x)\,, \qquad  
                        \widetilde{H}^q(x,\xi=0,t=0)\= \Delta q(x)\,,  
\ee
i.e.\ in the forward limit of zero momentum transfer and zero
skewness, two of them, namely $H$ and $\widetilde{H}$ reduce to
the usual unpolarized and polarized parton distributions (PDFs),
respectively. The other two GPDs, $E$ and $\widetilde{E}$, are not
accessible in DIS. Another property of the GPDs is the polynomiality which
comes about as a consequence of Lorentz covariance 
\be
\int^1_{-1}\, dx\, x^{n-1}\, H^q(x,\xi,t) \= \sum^{[n/2]}_{i=0}\,
h^q_{n,i}(t)\,\xi^i\,,
\label{pol}
\ee 
where $[n/2]$ denotes the largest integer smaller than or equal to
$n/2$. Eq.\ \req{pol} holds analogously for the other GPDs and, for
$n=1$ implies sum rules for the form factors of the nucleon, e.g.\
\be
F^q_1(t) \= h^q_{1,0}(t) \= \int^1_{-1}\, dx H^q(x,\xi,t)\,.
\label{sumrule}
\ee
Reinterpreting as usual a parton with a negative momentum fraction $x$
as a antiparton with positive $x$ ($H^{\bar{q}}(x)=-H^q(-x)$), one 
becomes aware that this sum rules provides the difference of the
contributions from quarks and antiquarks of given flavor to the
Dirac form factor of the nucleon. Introducing the combination
\be
H^q_v(x,\xi,t) \=   H^q(x,\xi,t) -  H^{\bar{q}}(x,\xi,t)\,,
\ee
which, in the forward limit, reduces to the usual valence quark density
$q_v(x)=q(x)-\bar{q}(x)$, one finds for the Dirac form factor the
representation
\be
F_1^{p(n)}(t)\= e_{u(d)} \int_0^1\, dx\, H_v^u(x,\xi,t) + e_{d(u)}
\int_0^1\, dx\, H_v^d(x,\xi,t)\,.
\label{pn-sr}
\ee     
Here, $e_q$ is the charge of the quark $q$ in units of the positron charge.
There might be contributions from other quarks, $s, c,...$, to the sum rule
\req{pn-sr}. These possible contributions are likely small ( in the
forward limit one has for instance $s(x) \simeq \bar{s}(x)$ \ci{CTEQ})
and neglected in the sum rule \req{pn-sr}. A representation anlogue to
\req{pn-sr} holds for the Pauli form factor replacing $H$ by $E$.

The isovector axial vector form factor, on the other hand, satisfies
the following sum rule
\be
F_A(t) \= \int_0^1 dx\, \Big[\widetilde{H}^u(x,\xi,t) -
                                  \widetilde{H}^d(x,\xi,t)\Big]
                   + 2 \int_0^1 dx\, \Big[\widetilde{H}^{\bar{u}}(x,\xi,t) -
                                  \widetilde{H}^{\bar{d}}(x,\xi,t)\Big]\,,
\label{axial-sr}
\ee
where $\widetilde{H}^{\bar{q}}(x)=\widetilde{H}^{q}(-x)$. At least for
small $t$ the magnitude of the second integral in Eq.\ \req{axial-sr}
reflects the size of the flavor-singlet combination 
$\Delta \bar{u}(x) - \Delta \bar{d}(x)$ of forward densities. This 
difference is poorly known, and at present there is no experimental 
evidence that it might be large \ci{HERMES}. In the analysis of the 
polarized PDFs performed by Bl\"umlein and B\"ottcher~Ref.\ \ci{BB} 
it is even zero. In a perhaps crude approximation the second term in 
\req{axial-sr} can be neglected.  

We also know how the GPDs evolve with the
scale. They satisfy positivity bounds and possess overlap
representations. But, I repeat, we don't know how to calculate them   
accurately from QCD at present. Thus, we have either to rely on models
or we have to extract the GPDs from experiment as it has been done for
the PDFs, see for instance Refs.\ \ci{CTEQ,BB}. The universality property
of the GPDs, i.e.\ their process independence, subsequently allows to
predict other hard exclusive reactions once the GPDs have been
determined in the analysis of a given process. This way QCD acquires a
predictive power for hard processes provided factorization holds. 

\section{Analysis of the zero-skewness GPDs}
At present the data basis is too sparse to allow for a phenomenological 
extraction of the GPDs as a function of the three variables $x$, $t$
and $\xi$. Lacking are in particular sufficient data on deeply virtual
Compton scattering. Available is, on the other hand, a fair amount of
nucleon form factor data, spread over a fairly large range of momentum
transfer, see references given in \ci{brash,DFJK4}. 
More form factor data will become available in the near future. 

The form factors represent the first moment of GPDs for any value of 
the skewness and, in particular, at $\xi=0$, see \req{pn-sr}, \req{axial-sr}. 
Mathematically one needs infinitely many moments to deduce the
integrand, i.e.\ the GPDs. However, from phenomenological experience
with particle physics one can expect the GPDs to be rather smooth
functions and, therefore, a small number of moments may suffer to fix
the GPDs. A first attempt in this direction adopts the extreme (and at
present the only feasible) point of view that the lowest moment of a
GPD suffices to determime it \ci{DFJK4} (see also Ref.\ \ci{guidal}). 
Indeed using recent results on PDFs \ci{CTEQ,BB} and form factor data
\ci{brash} as well as suitable, physically motivated parameterizations 
of the GPDs with a few parameters adjusted to data, one can indeed
carry through this analysis. Needless to say that this method while 
phenomenologically succesful as I will discuss below, does not lead to
a unique result. Other parameterizations which may imply different
physics, cannot be excluded in the present stage of the GPD
analysis. In principle this can be remedied by the inclusion of higher 
order moments from lattice QCD. Provided the lattice results are
obtained in a scenario with light quarks or reliably extrapolated to
the chiral limit, a combined analysis of form factor and lattice data
will lead to improved results on the GPDs with a lesser or even no
dependence on the chosen parameterization. The analysis advocated for 
in Ref.\ \ci{DFJK4} can easily accomodate higher order moments. The
LHPC \ci{SESAM} and QCDSF \ci{QCDSF} collaborations have
recently presented results on GPD moments in scenarios with pion 
masses around $800\,\mev$. These results have not used in the 
analysis presented in Ref.\ \ci{DFJK4} for obvious reasons but they 
indicate that, in a few years, the quality of the lattice results may 
suffice for use in a GPD analysis.        

In Ref.\ \ci{DFJK4} $\xi=0$ is chosen (implying $x_1=x_2$) and a
parameterization of the GPDs is exploited that combines the usual PDFs
with an exponential $t$ dependence (the argument $\xi=0$ is dropped in  
the following for convenience)
\be
H_v^q(x,t) \= q_v(x) \exp[tf_q(x)]\,,
\label{ansatz}
\ee
where the profile function reads 
\be
f_q(x) \= \big[\alpha'\, \log(1/x) + B_q\big](1-x)^{n+1} + A_q x (1-x)^n\,.
\label{profile}
\ee 
This ansatz is motivated by the expected Regge behaviour at low $t$
and low $x$ \ci{arbarbanel} (where $\alpha'$ is the Regge slope for
which the value $0.9\,\gev^2$ is imposed). For large $t$ and large
$x$, on the other hand, one expects a behaviour like $f_q \sim 1-x$
from the overlap model \ci{barone,DFJK1,DFJK3,rad98}. The ansatz \req{ansatz},
\req{profile} interpolates between the two limits smoothly~\footnote{
The parameter $B_q$ is not needed if $\alpha'$ is freed. A value of
$\simeq 1.4$ for $\alpha'$ leads to a fit to the data of about the
same quality and with practically the same results for the GPDs.} 
and allows the generalization to the case $n=2$. 
The ansatz \req{ansatz}, \req{profile} matches the following criteria
for a reasonable parameterization
\begin{itemize}
\item simplicity
\item consistency with theoretical and phenomenological constraints
\item plausible interpretation of the parameters (if possible)
\item stability with respect to variations of PDFs
\item stability under evolution (scale dependence of GPDs can be
  absorbed into parameters)
\end{itemize} 

A fit of the ansatz \req{ansatz}, \req{profile} to the data on the
Dirac form factor ranging from $t=0$ up to $\simeq 30\, \gev^2$,
exploiting the sum rule \req{pn-sr} and using the CTEQ PDFs \ci{CTEQ}, 
leads to very good results with the three parameters 
\be
B_u\=B_d\=(0.59\pm 0.03)\,\gev^{-2}\,, \quad A_u\=(1.22\pm 0.020)\,\gev^{-2} 
                                       \quad A_d\= (2.59\pm 0.29)\,\gev^{-2}\,, 
\ee
quoted for the case $n=2$ and at a scale of $\mu=2\,\gev$.

\begin{figure}
\begin{center}
\includegraphics[width=.35\textwidth, height=.35\textwidth,
  bb=77 448 399 786,clip=true] {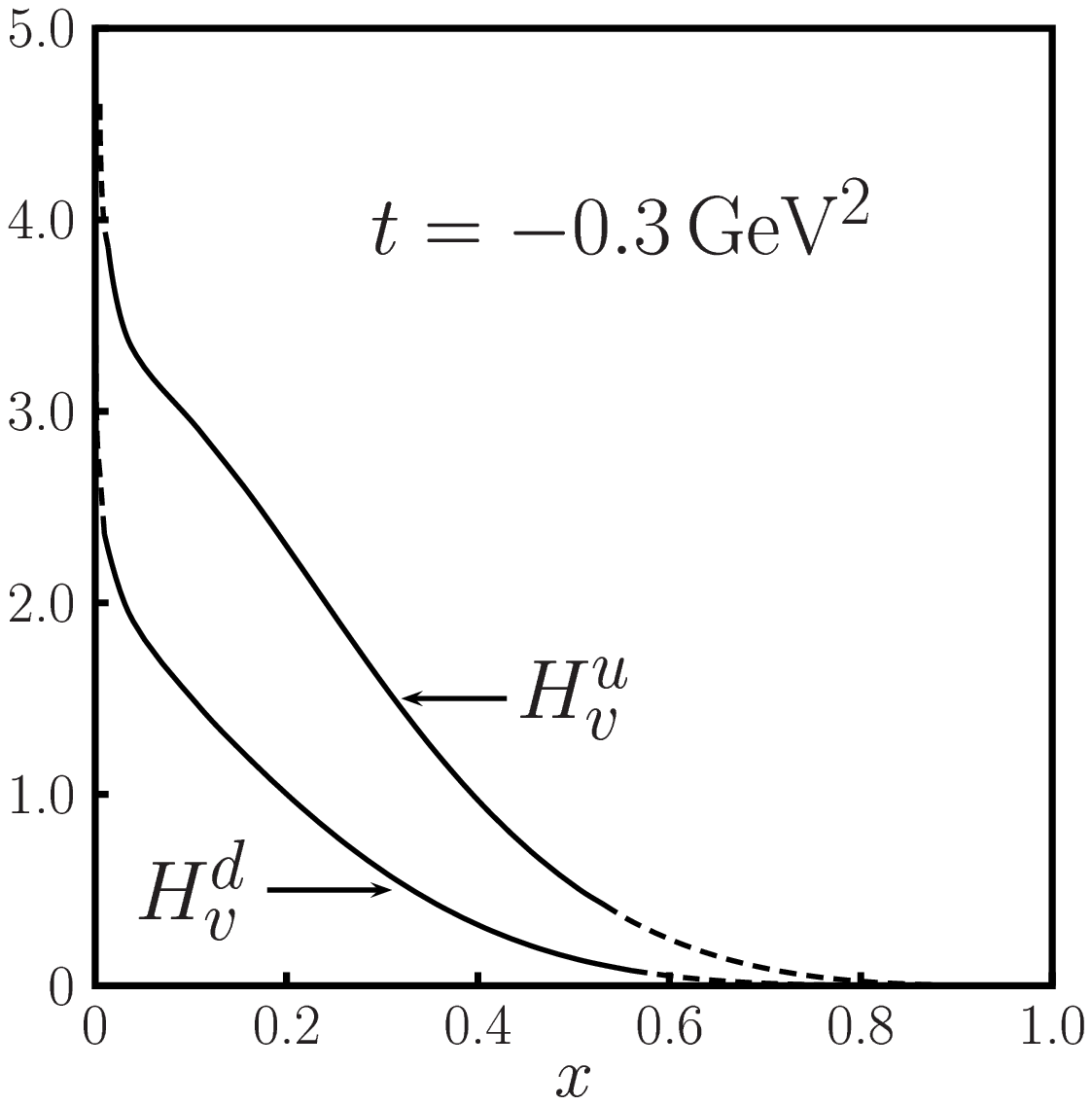}
\hspace{4em} 
\includegraphics[width=.35\textwidth, height=.34\textwidth,
  bb=68 364 399 683,clip=true] {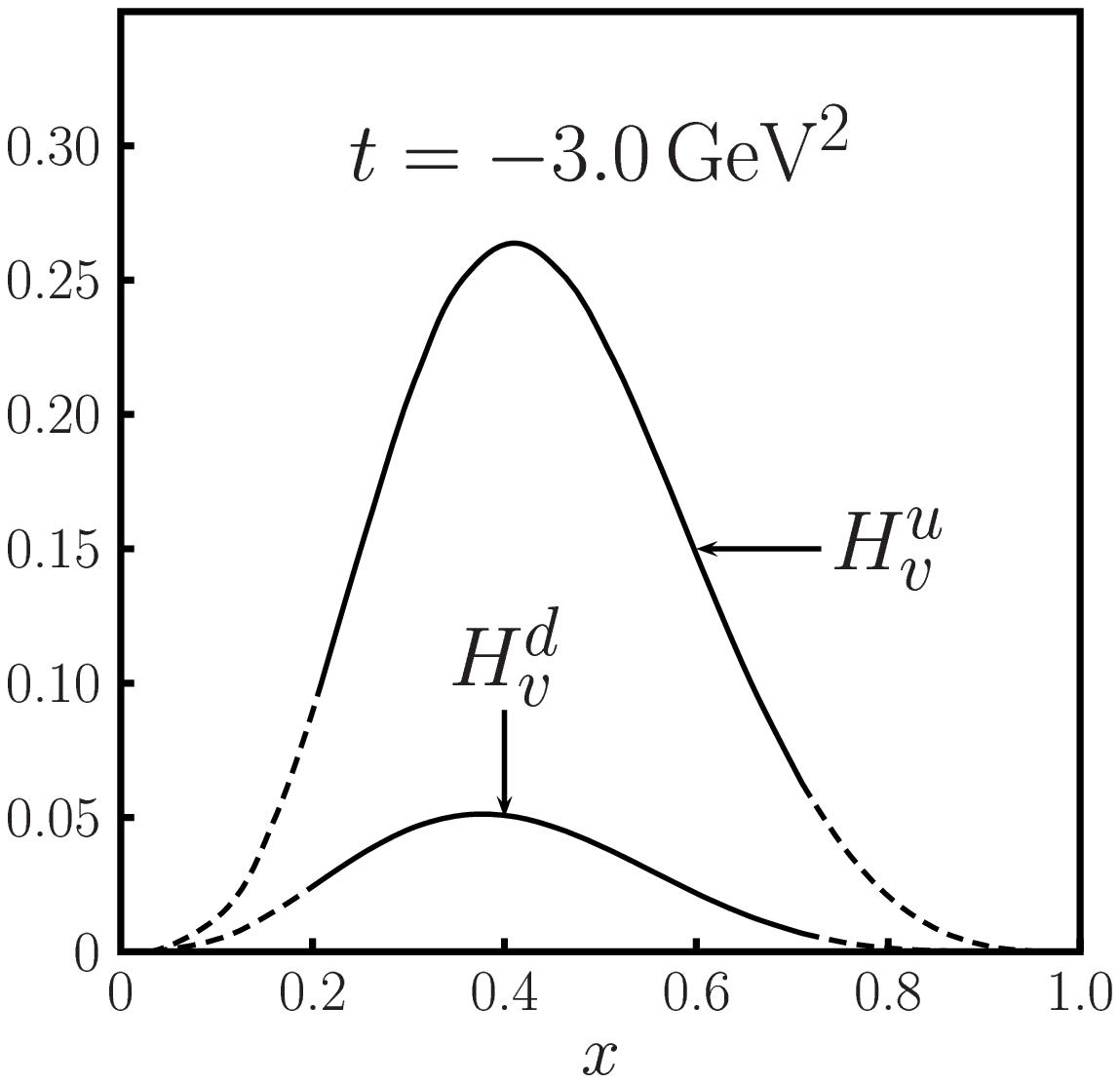}
\end{center}
\vspace*{-1.6em}
\caption{\label{fig:Hgpd} Result for the valence GPD $H_v^q(x,t)$ at
the scale $\mu=2 \gev$ and for $n=2$ obtained in the analysis
presented in \cite{DFJK4}. The dashed lines indicate regions of $x$
where only $5\%$ of the sum rule \req{pn-sr} is accumulated.}
\label{fig:gpdH}
\end{figure}

In Fig.\ \ref{fig:gpdH} the results for $H$ are shown at two values
of $t$. While at small $t$ the behaviour of the GPD still reflects
that of the parton densities it exhibits a pronounced maximum at
larger values of $t$. The maximum moves towards $x=1$ with increasing
$t$ and becomes more pronounced. In other words only a limited range of
$x$ contributes to the form factor substantially and this range moves
with $t$ in parallel with the position of the maximum of $H$. 
The quality of the fit is very similar in both the cases, $n=1$ and 2;
the results for the GPDs and related quantities agree very well with
each other. Substantial differences between the two results only occur
for very low and very large values of $x$, i.e.\ in  regions which are
nearly insensitive to the form factor data. It is the physical
interpretation of the results which favours the fit with $n=2$. Indeed
the average distance between the struck quark and the cluster of spectators
becomes unphysical large for $x\to 1$ in the case $n=1$; it grows like
$\sim (1-x)^{-1}$ while, for $n=2$, it tends to a constant value of
about 0.5 fm \ci{DFJK4}. 

The analogue analysis of the axial and Pauli form factors, with
parameterizations similar to Eqs.\ \req{ansatz}, \req{profile}, provides the
GPDs $\widetilde{H}$ and $E$. They behave similar to $H$. Noteworthy
differences are the opposite signs of $\widetilde{H}^u$ ($E^u$) and
$\widetilde{H}^d$ ($E^q$) and the approximately same magnitude of 
$E^u$ and $E^d$ at least for not too large values of $t$. For $H^q$
and $\widetilde{H}^q$, on the other hand, the $d$-quark contributions
are substantially smaller in magnitude than the $u$-quark ones, see
Fig.~\ref{fig:gpdH}. Since there is no data available for the
pseudoscalar form factor of the nucleon the GPD $\widetilde{E}$ cannot
be determined this way. 

I repeat the results for the GPDs are not unique. An alternative
ansatz is for instance 
\be
H_v^q(x,t) \= q_v(x)\Big[1-t f_q(x)/p\Big]^{-p}\,.
\label{ansatz-power}
\ee
Although reasonable fits to the form factors are obtained with it for
$p\gsim 2.5$ it is physically less appealing than \req{ansatz}: the
combination of Regge behaviour at small $x$ and $t$ with the dynamics
of the Feynman mechanism is lost. The resulting GPDs have a broader
shape and $H(x=0,t)$ remains finite. Thus, small $x$ also contribute
to the high-$t$ form factors for the ansatz \req{ansatz-power}.

\section{Moments and interpretation}
\begin{figure}
\begin{center}
\includegraphics[height=.35\textwidth,
 bb = 48 296 444 655,clip=true]{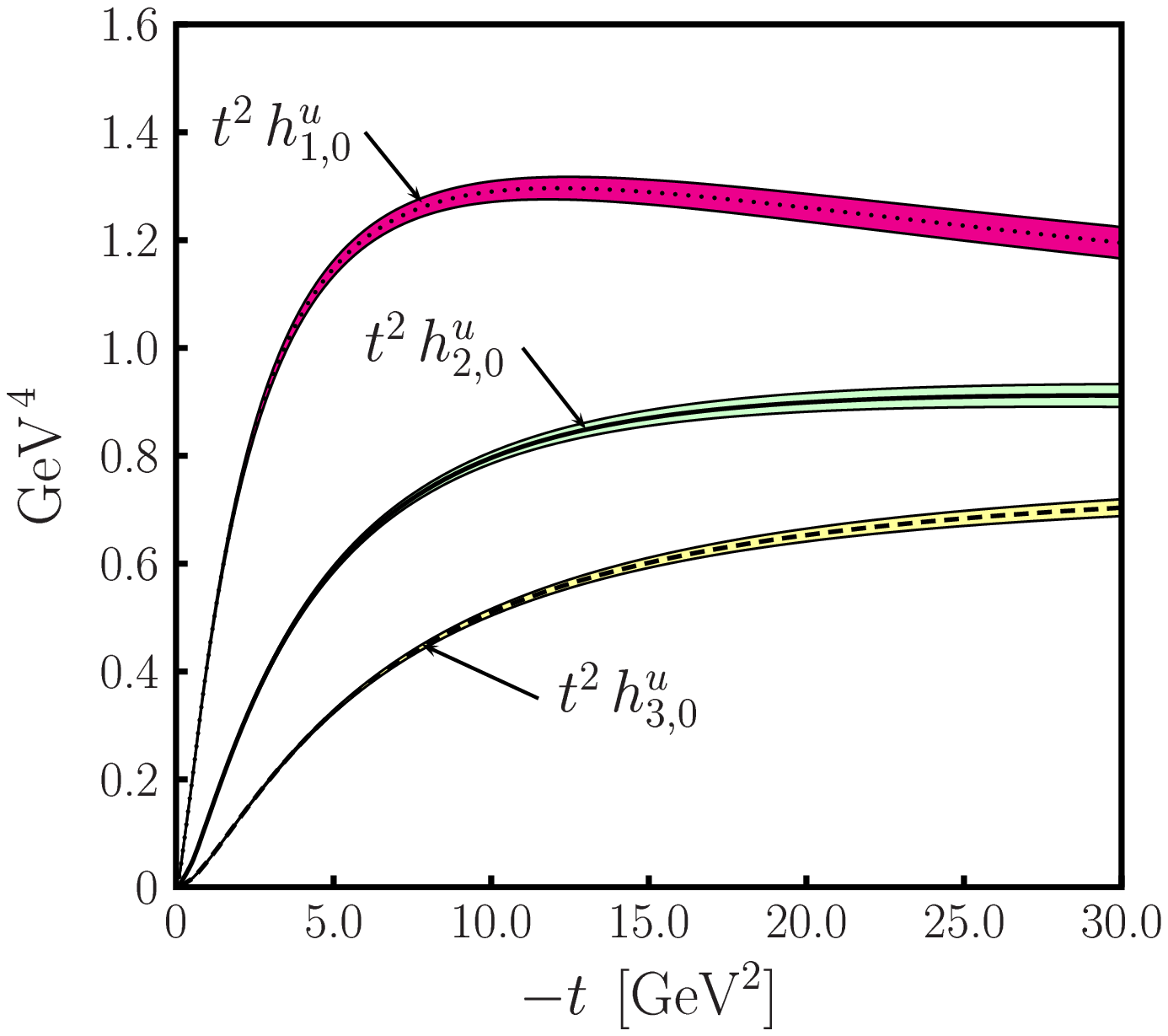}
\hspace{4em} 
\includegraphics[height=.35\textwidth,
 bb = 104 341 467 686,clip=true]{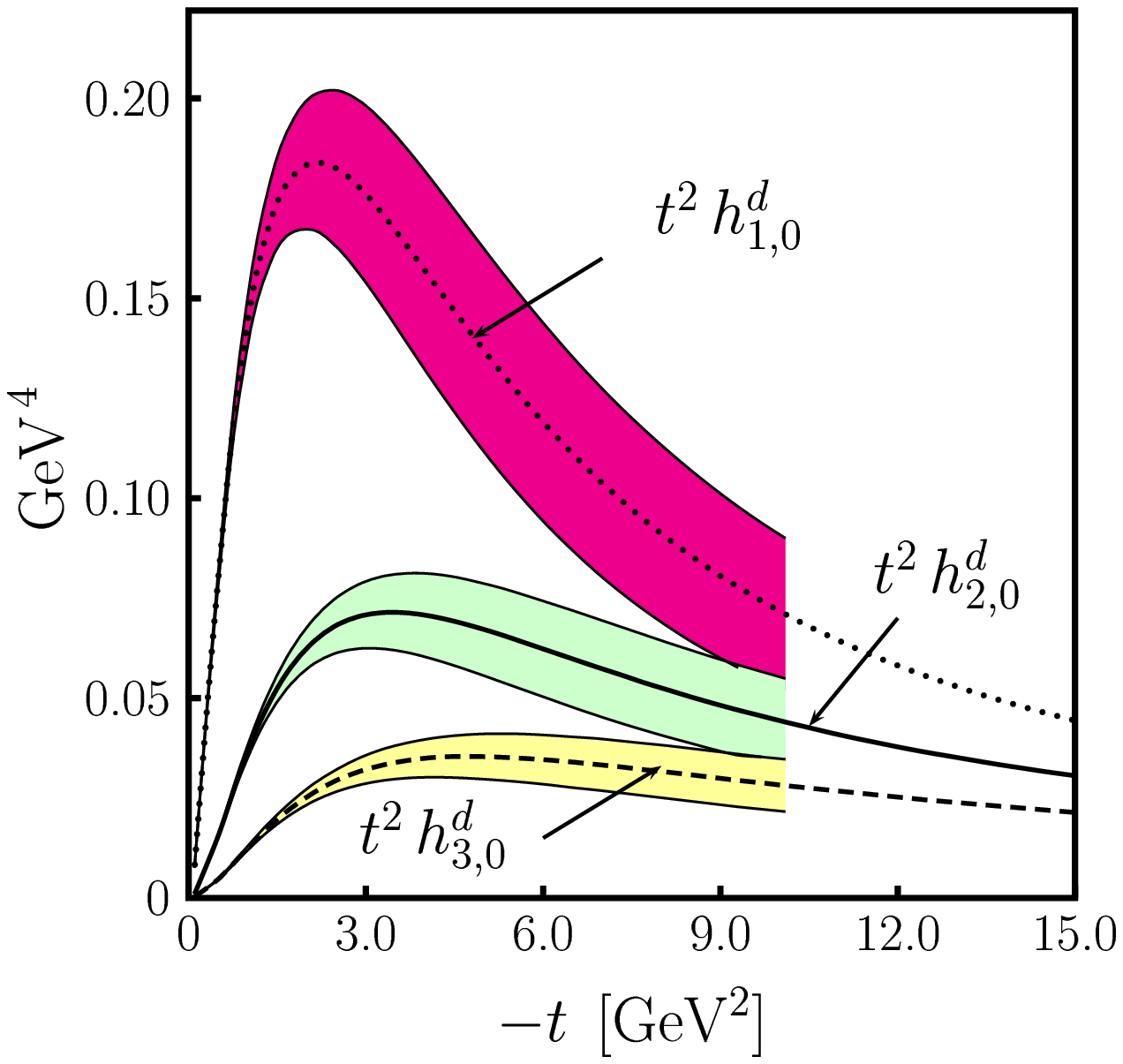}
\end{center}
\vspace*{-1.6em}
\caption{\label{fig:moments} The first three moments of valence GPDs
$H_v^u$ (left) and $H_v^d$ (right), scaled with $t^2$.  The error
bands denote the parametric uncertainty resulting from the fit to the
Dirac form factors $F_1^{p}$ and~$F_1^{n}$.}
\end{figure}

Having the zero-skewness GPDs at disposal one can evaluate various
moments, some of them are displayed in Fig.\ \ref{fig:moments}.  Comparison
with recent results from lattice QCD \ci{SESAM,QCDSF} reveals 
remarkable agreement of their $t$ dependencies given the uncertainties 
in the GPD analysis \ci{DFJK4} and in the lattice calculations \ci{SESAM,QCDSF}.
An interesting property of the moments is that the $u$ and $d$ quark
contributions drop with different powers of $t$. These powers are
determined by the large-$x$ behaviour of $q_v$ and $f_q$, namely
\be
q_v \sim (1-x)^{\beta_q}\,, \qquad f_q \sim (1-x)^2\,,
\ee
where $\beta_u\simeq 3.4$ and $\beta_d\simeq 5$ \ci{CTEQ}. 
Because of the pronounced maximum the GPDs exhibits, see Fig.\
\ref{fig:gpdH}, the sum rule \req{pn-sr} can be evaluated in the
saddle point approximation. For large t this leads to
\be
h^q_{1,0} \sim |t|^{-(1+\beta_q)/n}\,.
\label{power}
\ee
With the CTEQ values for $\beta_q$ one obtains a drop of the form
factor $F_{1v}^u$ slightly faster than $t^{-2}$ while the $d$-quark
form factor falls as $t^{-3}$. Strengthened by the charge factor the
$u$-quark contribution dominates the proton's Dirac form factor for
$t$ larger than about $5\,\gev^2$, the $d$-quark contribution amounts to
less than $10\%$. High quality neutron form factor data above
$3\,\gev^2$ would allow for a direct examination of the different powers.
The power behaviour bears resemblance to the Drell-Yan relation
\ci{DY}. In fact the common underlying dynamics is the Feynman
mechanism~\footnote{
The Feynman mechanism applies in the soft region where $1-x\sim
\Lambda/\sqrt{-t}$ and the virtualities of the active partons are
$\sim \Lambda\sqrt{-t}$ ($\Lambda$ is a typical hadronic scale).}.
The Drell-Yan relation is, however, an asymptotic result 
($x\to 1$, $t\to \infty$) which bases on the assumption of valence
Fock state dominance. In the GPD analysis Eq.\ \req{power} holds 
provided the saddle point lies in the region where the bulk of the 
contribution to the Dirac form factor is accumulated.

A combination of the second moments of $H$ and $E$ at
$t=0$ is Ji's sum rule \cite{ji97} which allows for an evaluation of 
the valence quark contribution to the orbital angular momentum the quarks 
inside the proton carry
\be
\langle L_v^q\rangle \= \frac12\,\int_0^1 dx \Big[xE_v^q(x,t=0) + x q_v(x)-
  \Delta q_v(x)\Big]\,.
\ee
A value of -0.08 has been found in Ref.\ \cite{DFJK4} at the scale
$\mu=2\,\gev$ for the average
valence quark contribution to the orbital angular momentum. 

\begin{figure}[t]
\begin{center}
\includegraphics[width=0.35\textwidth,bb=115 207 485 660, angle=-90,clip=true]
{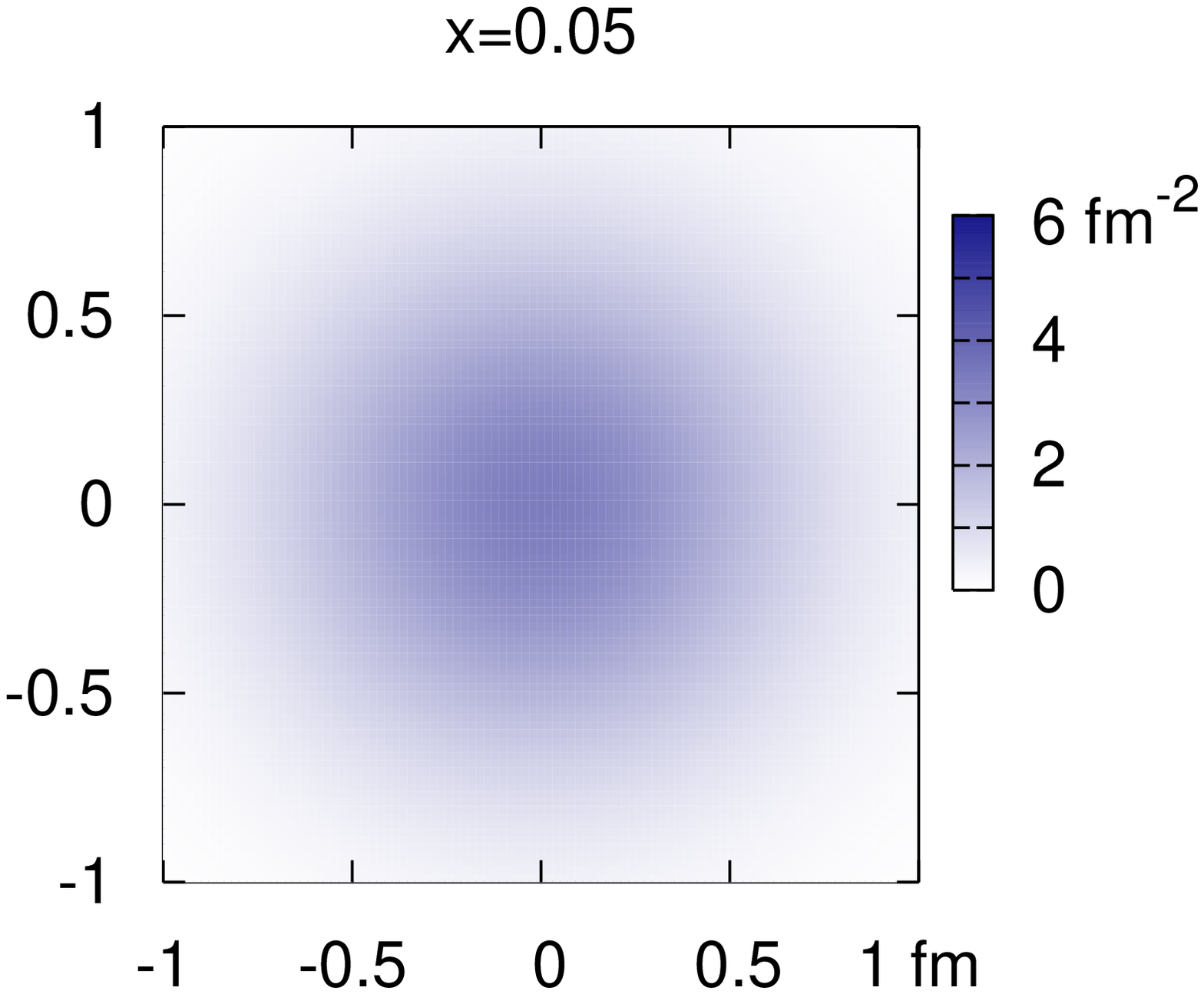}
\includegraphics[width=0.35\textwidth,bb=115 207 485 660,angle=-90,clip=true]
{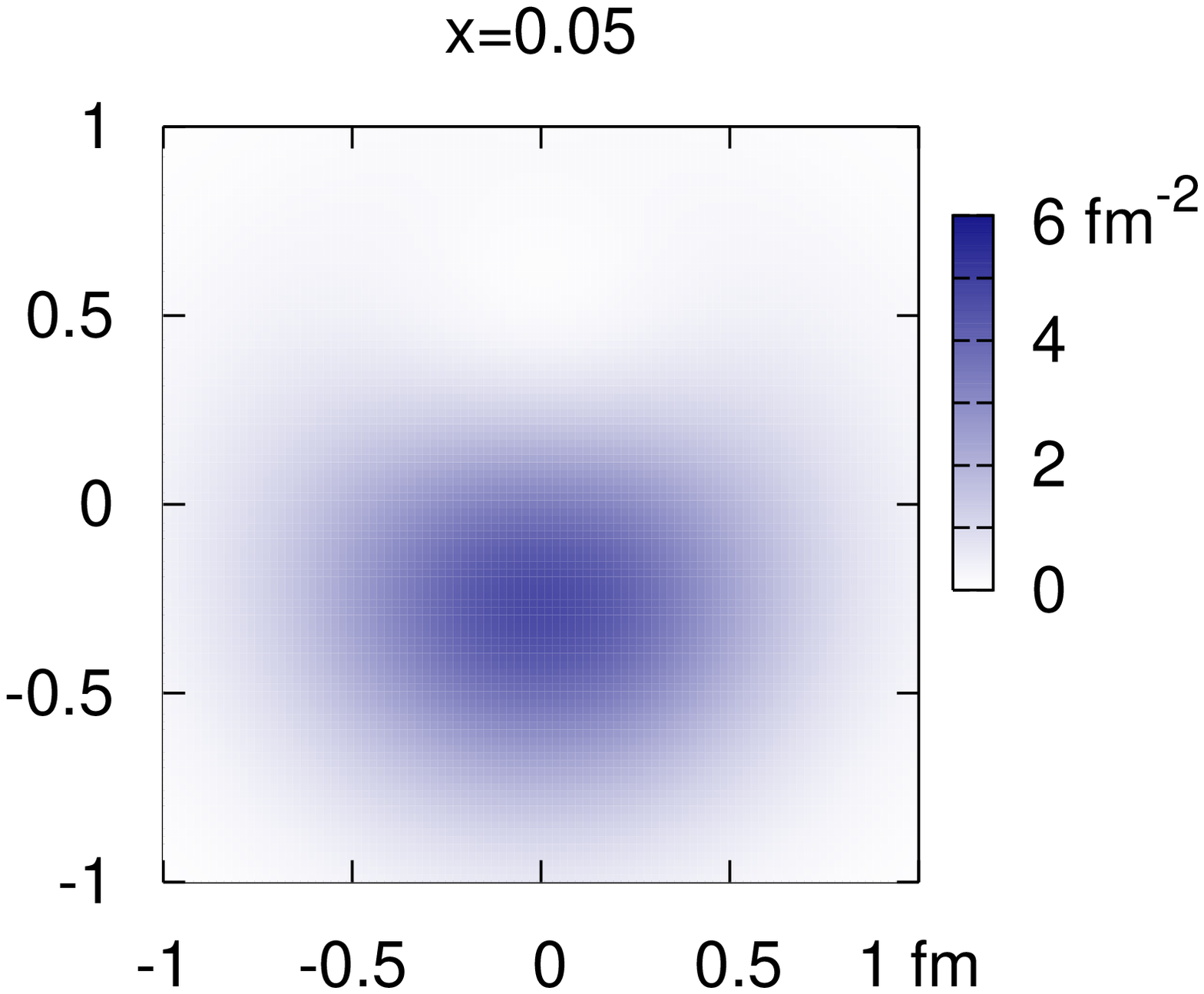}
\\
\includegraphics[width=0.35\textwidth,bb=115 207 485 660,angle=-90,clip=true]
{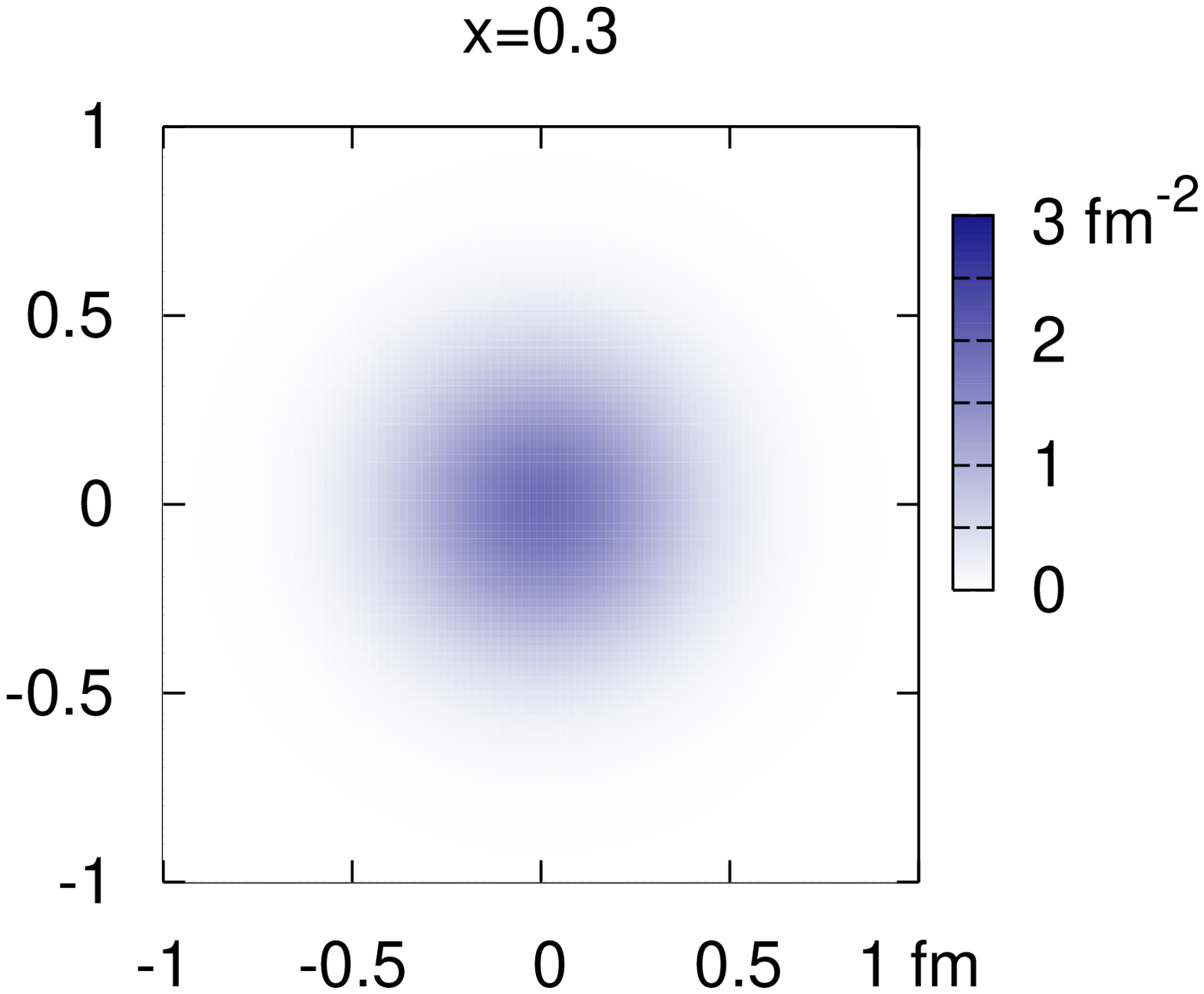}
\includegraphics[width=0.35\textwidth,bb=115 207 485 660,angle=-90,clip=true]
{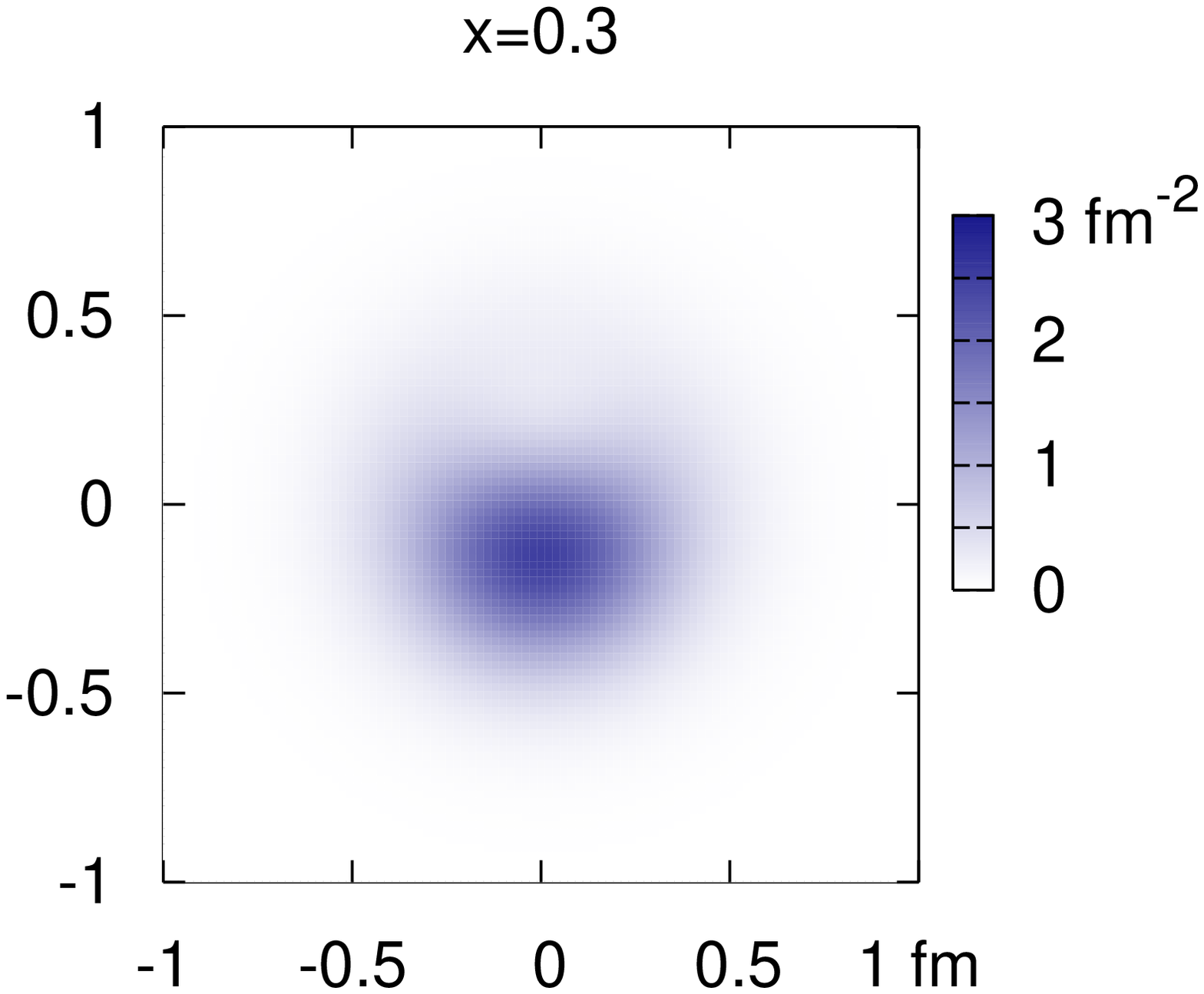}
\vspace{-4pt}
\end{center}
\caption{\label{fig:tomo_d} Tomography plots of $d_v(x,{\bf b})$
(left) and $d_v^X(x,{\bf b})$ (right) in the transverse $b^X$--$b^Y$
plane (in units of fm).  The longitudinal momentum fraction $x$ is
fixed to $0.05$ (top) and $0.3$ (bottom). 
The scale of intensity is given on the right hand
side of the plots.}
\end{figure}

In contrast to the parton distributions which only provide
information on the longitudinal distribution of quarks inside the
nucleon, GPDs also give access to the transverse structure of the
nucleon by Fourier transforming the GPD with respect to
$\sqrt{-t}$. As shown by Burkardt \ci{burk}, a density interpretation
of the zero-skewness GPDs is obtained in the mixed representation of
longitudinal and transverse position in the infinite-momentum
frame. In particular
\be
q_v(x,{\bf b}) \= \int \frac{d^2{\bf \Delta}}{(2\pi)^2}\, 
                e^{-i{\bf b}\cdot{\bf \Delta}}\, H_v^q(x,t=-\Delta^2)\,,
\ee  
gives the probability to find a valence quark with longitudinal momentum
fraction $x$ and impact parameter ${\bf b}$. Together with the
analogue Fourier transform of $E_v^q(x,t)$ one can form the
combination ($m$ being the mass of the proton)
\be
q_v^X(x,{\bf b}) \= q_v(x,{\bf b}) - \frac{\;b^Y}{m}\,
\frac{\partial}{\partial {\bf b}^2}\, e_v^q(x,{\bf b})\,,
\ee
which gives the probability to find an unpolarized valence quark with
momentum fraction $x$ and impact parameter ${\bf b}=(b^X,b^Y)$ in a
proton that moves rapidly along the $Z$ direction and is polarized
along the $X$ direction \ci{burk}. In Fig.\ \ref{fig:tomo_d} the
results for the GPDs are shown as tomography plots in the impact parameter
space for fixed momentum fractions. For small $x$ one observes a very 
broad distribution while, at large $x$, it becomes more focussed 
on the center of momentum defined by $\sum_i x_i {\bf b}_i=0$ 
($\sum_i x_i=1$). In a proton that is polarized in the $X$ direction the 
symmetry around the $Z$ axis is lost and the center of the density is 
shifted in the $Y$ direction away from the center of momentum,
downward for $d$ quarks and upward for $u$ ones. Thus, a polarization 
of the proton induces a flavor segregation in the direction orthogonal 
to the direction of the polarization and the proton momentum.        

\section{Applications: Compton scattering and photoproduction}
\begin{figure}[t]
\begin{center}
\includegraphics[width=0.35\textwidth,height=0.3\textwidth, 
bbllx=120pt,bblly=560pt,bburx=280pt,bbury=680pt,clip=true]{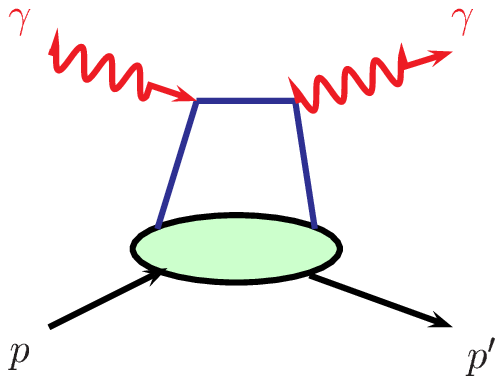} 
\includegraphics[width=.35\textwidth, height=.35\textwidth,
  bb= 50 107 387 430,clip=true]{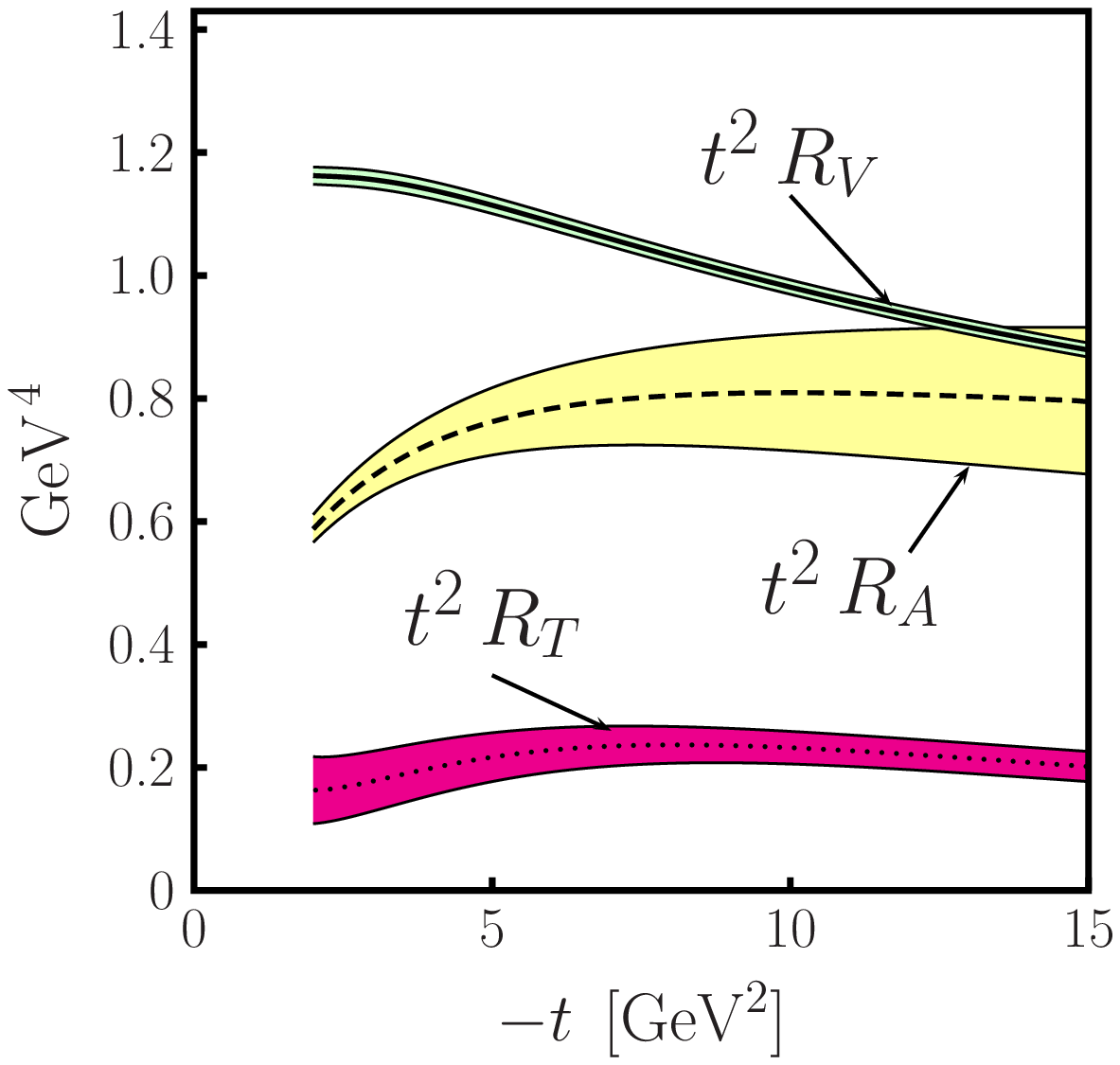}
\end{center}
\vspace*{-1.6em}
\caption{{}Handbag diagram for Compton scattering (left) and the
Compton form factors, scaled by $t^2$, evaluated from the GPDs
determined in \protect\ci{DFJK4} (right). The bands represent the
parametric uncertainties of the form factors~\protect\ci{DFJK4}.} 
\label{fig:handbag}
\end{figure}

As I discussed in the preceeding sections the analysis of the GPDs
gives insight in the transverse distribution of quarks inside the
proton. However, there is more in it. With the $\xi=0$ GPDs at
hand one can now predict hard wide-angle exclusive reactions like
Compton scattering off protons or meson photo- and electroproduction~\footnote{
For deep virtual exclusive processes where the virtuality
of the photon is large while the momentum transfer from the initial to
the final proton is small,  skewness is fixed by Bjorken-$x$
($\xi\simeq x_{Bj}/(2-x_{Bj})$). Hence, the GPDs for non-zero skewness are
required in calculations of such processes.}. 
For these reactions one can work in a symmetrical frame where skewness 
is zero. A symmetrical frame is, for instance, a c.m.s. rotated in such a way
that the initial and final protons have the same light-cone plus
component. Hence, $\xi=0$. It has been argued \ci{DFJK1,rad98,hanwen} 
that, for large Mandelstam variables ($s, \;-t, \; -u \gg m^2$),
the amplitudes for these processes factorize in a hard partonic
subprocess, e.g.\ Compton scattering off quarks - see Fig.\
\ref{fig:handbag}, and in form factors representing $1/x$-moments of
zero-skewness GPDS. For Compton scattering these form factors read 
\bea
R_V(t) &\simeq&\sum_q e_q^2 \int_{0}^1 \frac{dx}{x}\, H^q_v(x,t)\,, \nn\\
R_A(t) &\simeq& \sum_q e_q^2 \int_{0}^1 \frac{dx}{x}\, 
\widetilde{H}^q_v(x,t)\,, \nn\\[0.2em]
R_T(t) &\simeq&\sum_q e_q^2 \int_{0}^1 \frac{dx}{x}\, E^q_v(x,t)\,.
\label{Compton-formfactors}
\eea
These relations only hold approximately even though with a very high degree
of accuracy at large $t$ since contributions from sea quarks can safely
be neglected~\footnote{
A rough estimate of the sea quark contribution may be obtained by
using the ansatz \req{ansatz} with the same profile function for
valence and sea quarks but replacing the valence quark density with
the CTEQ \ci{CTEQ} antiquark ones.}.
A pseudoscalar form factor related to the GPD $\widetilde{E}$,
decouples in the symmentric frame. Numerical results for the Compton
form factors are shown in Fig.\ \ref{fig:handbag}. 
Approximately the form factors $R_i$ behave $\sim t^{-2}$.
The particular flat behaviour of the scaled form factor $t^2R_T$ is a
consequence of a cancellation between the $u$ and $d$-quark
contributions. The ratio $R_T/R_V$ behaves differently from the 
corresponding ratio of their electromagnetic analogues $F_2$ and $F_1$. 

The handbag contribution leads to the following leading-order 
result for the Compton cross section \ci{DFJK1,HKM} 
\bea
\frac{d\sigma}{dt} &=& \frac{d\hat{\sigma}}{dt} \left\{ \frac12\, \Big[
R_V^2(t)\, + \frac{-t}{4m^2} R_T^2(t) + R_A^2(t)\Big] \right.\nn\\
&&\hspace*{-0.5cm}\left.  - \frac{u s}{s^2+u^2}\,
\Big[R_V^2(t)\,+ \frac{-t}{4m^2} R_T^2(t) - R_A^2(t)\Big]\right\}\,,
\label{dsdt}
\eea
where $d\hat{\sigma}/dt$ is the Klein-Nishina cross section for
Compton scattering off massless, point-like spin-1/2 particles of
charge unity. Next-to-leading order QCD corrections to the subprocess
have been calculated in Ref.\ \ci{HKM}. They are not displayed in 
\req{dsdt} but taken into account in the numerical results discussed
below. Another interesting observable in Compton scattering is the
helicity correlation, $A_{LL}$,  between the initial state photon and
proton or, equivalently, the helicity transfer, $K_{LL}$, from the
incoming photon to the outgoing proton. In the handbag approach one
obtains 
\cite{HKM} 
\be
A_{LL}\=K_{LL}\simeq \frac{s^2 - u^2}{s^2 + u^2}\, 
                    \frac{R_A(t)}{R_V(t)} + O(\alpha_s)\,,
\label{all}
\ee  
where the factor in front of the form factors is the corresponding
observable for $\gamma q\to \gamma q$. The result \req{all} is a
robust prediction of the handbag mechanism, the magnitude of the
subprocess helicity correlation, $\hat{A}_{LL}$, is only diluted
somewhat by the ratio of the form factors $R_A$ and $R_V$. It is to be
stressed that $A_{LL}$ and $K_{LL}$ are identically in the handbag
approach because the quarks are assumed to be massless and consequently 
there is no quark helicity flip. For an alternative approach, see 
Ref.\ \cite{miller}. It is also important to keep in mind that both
the results, \req{dsdt} and \req{all}, hold for $s,\;-t,\;-u\gg m^2$.

\begin{figure}[t]
\begin{center}
\includegraphics[width=.48\textwidth, height=0.43\textwidth,
  bb= 153 264 572 715,clip=true]{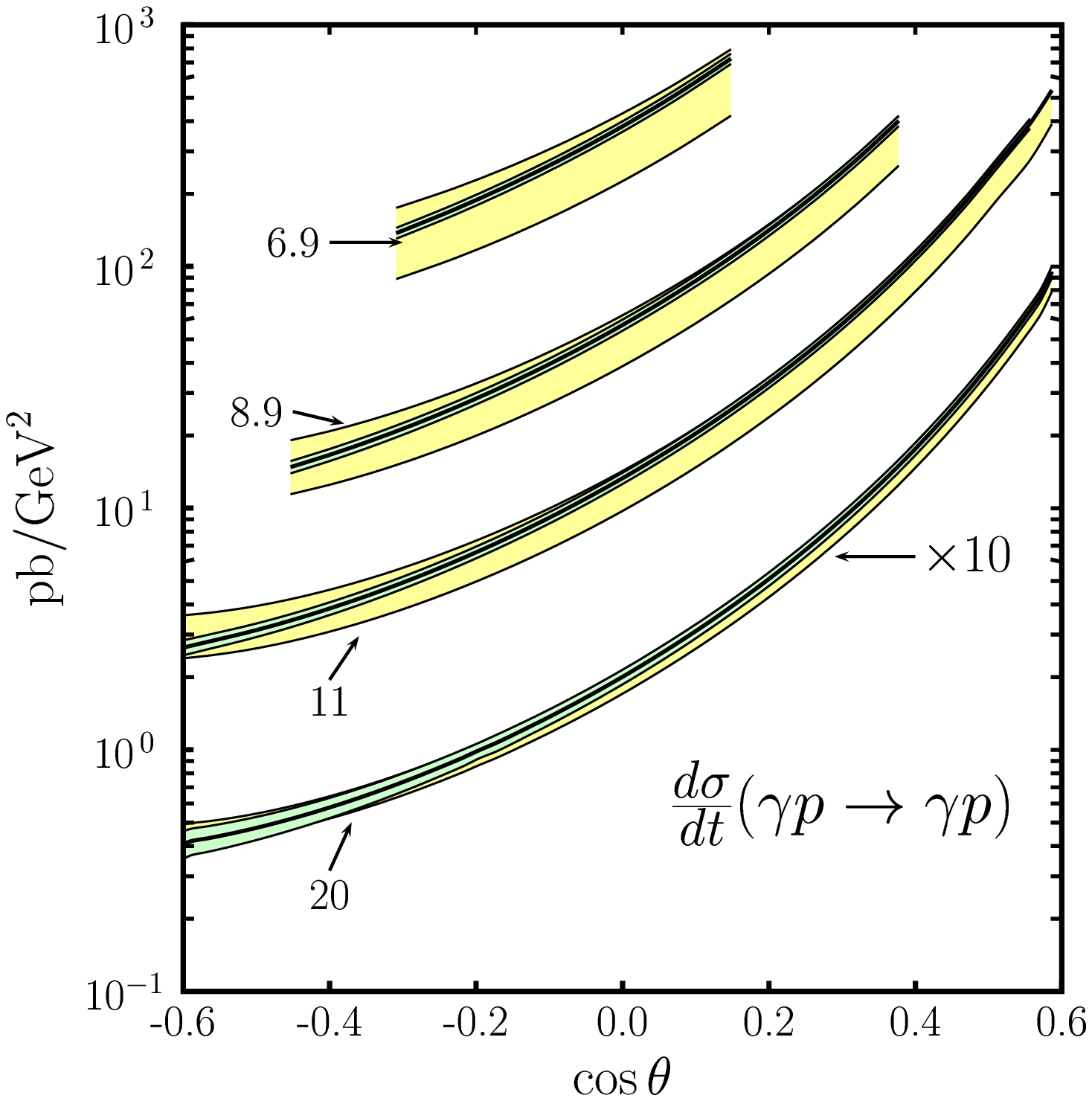}
\hspace{4em}
\includegraphics[width=.40\textwidth, 
  bb= 122 340 433 684,clip=true]{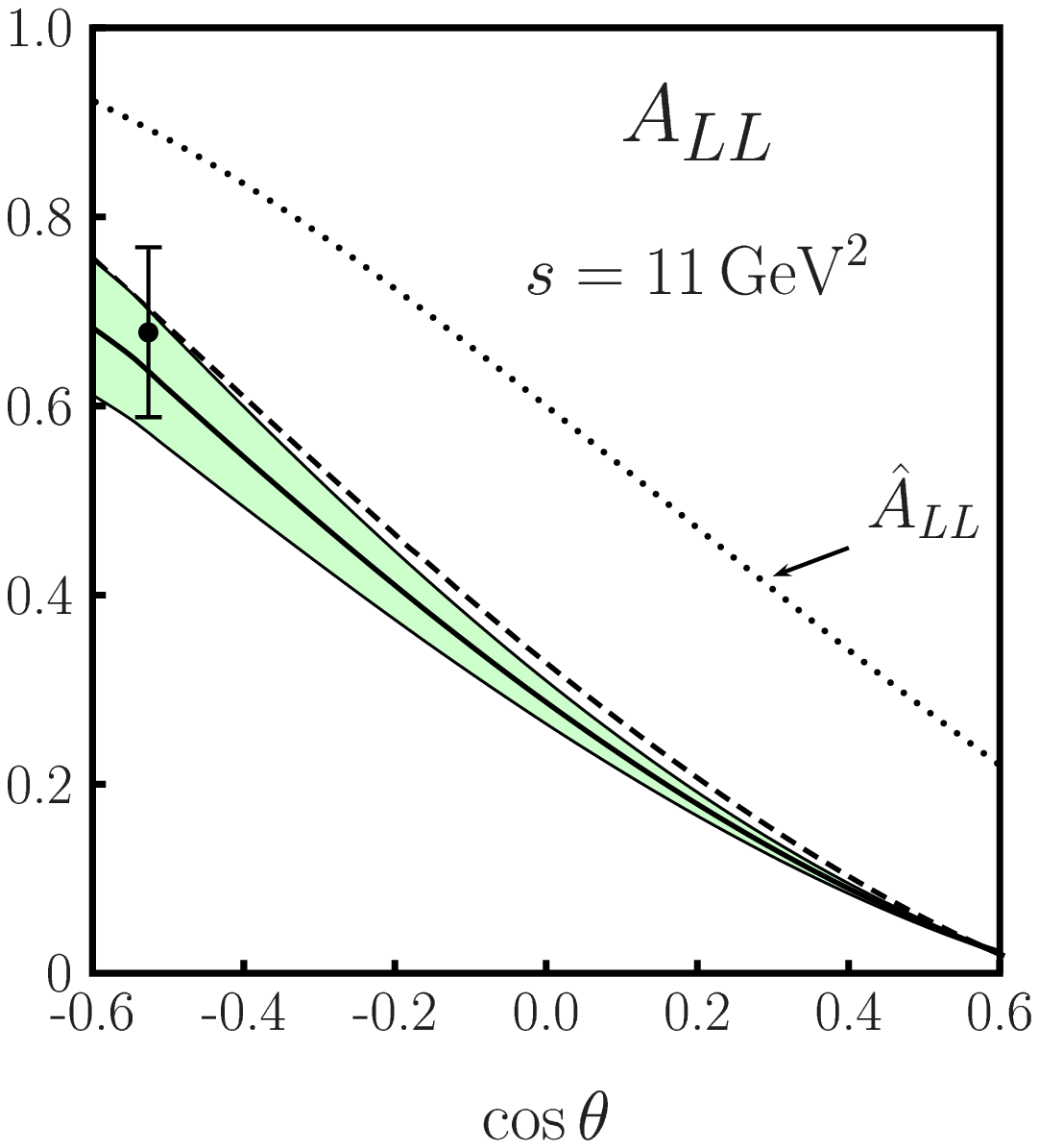}
\end{center}
\vspace*{-1.6em}
\caption{\label{fig:wacs-obs} The unpolarized cross section (left) and
the helicity correlation parameter $A_{LL}$ (right) for wide-angle
Compton scattering as functions of the c.m.s.\ scattering angle
$\theta$. The cross section is shown at $s= 6.9, 8.9, 11$ and
$20\gev^2$, the latter cross section is multiplied by 10.  
The helicity correlation is presented at $s=11\,\gev^2$ (solid line 
with error band). The dashed line is the prediciton at 
$s=20\,\gev^2$ and $\hat{A}_{LL}$ is the result for Compton scattering 
off quarks. The data point \protect\ci{hamilton} is measured at 
$s=6.9\,\gev^2$. Both observables are evaluated at NLO QCD with the 
Compton form factors shown in Fig.~\protect\ref{fig:handbag}. The
error bands are explained in the text.}
\end{figure}

Inserting the Compton form factors \req{Compton-formfactors} into 
Eqs.\ \req{dsdt} and \req{all}, one can predict the Compton cross
section in the wide-angle region as well as the helicity correlation 
$A_{LL}=K_{LL}$. The results for sample values of $s$ are shown in Fig.\ 
\ref{fig:wacs-obs}. The inner bands of the predictions for $d\sigma/dt$
reflect the parametric errors of the form factors, essentially that 
of the vector form factor which dominates the cross section. 
The outer bands indicate estimates of the target mass corrections, 
see \cite{DFHK}. As a minimum condition to the kinematical 
approximations made in the handbag approach, predictions are only
shown for $-t$ and $-u$ larger than about $2.4\, \gev^2$. The JLab 
E99-114 collaboration~\cite{nathan} will provide accurate cross
section data soon which will allow for a crucial examination of the 
predictions from the handbag mechanism.

The JLab Hall A collaboration \cite{hamilton} has presented a first
measurement of $K_{LL}$ at $s=6.9\,\gev^2$ and $t=-4\,\gev^2$.
The kinematical requirement of the handbag mechanism, $s,\; -t,\; -u
\gg m^2$, is not satisfied for this measurement since $-u$ is only
$1.13\,\gev^2$. One therefore has to be very cautious when comparing
this experimental result with the handbag predictions, there might be 
large dynamical and kinematical corrections. Nevertheless the
agreement of this data point with the prediction
from the handbag at $s=11\,\gev^2$ is, with regard to the mild energy
dependence of $K_{LL}$ in \req{all}, a non-trivial and promising fact.  
Polarization data at higher energies are desired as well as a
measurement of the angle dependence. The Jlab Hall A collaboration
\ci{hamilton} has also measured the polarization transfer $K_{LS}$ from a
longitudinally polarized incident photon to the sideway polarization of
the outgoing proton. In Ref.\ \ci{hamilton} a value of $0.114\pm 0.078 \pm 0.04$
is obtained for it at the same kinematics as for $K_{LL}$ while we
found $0.10 \pm 0.02$ at this admittedly small energy.  

The handbag approach also applies to wide-angle photo- and
electroproduction of pseudoscalar and vector mesons.
The amplitudes again factorize into a parton-level subprocess, $\gamma
q\to M q$, and form factors which represent $1/x$-moments of GPDs
\cite{hanwen}. Their flavor decomposition differs from those appearing
in Compton scattering, see \req{Compton-formfactors}. Here, it
reflects the valence 
quark structure of the produced meson. Since the GPDs and, hence, the
form factors for a given flavor, $R_i^q$, $i=V,A,T$ are process
independent they are known from the analysis of Ref.~\cite{DFJK4} for
$u$ and $d$ quarks (if the contributions from sea quarks can be ignored). 
Therefore, the soft physics input to calculations of photo-and
electroproduction of pions and $\rho$ mesons within the handbag
approach is now known.

One may also consider the time-like process
$\gamma\gamma\leftrightarrow p\bar{p}$ in the handbag approach
\ci{DKV}. Similar form factors as in the space-like region occur which
are now functions of $s$ and represent moments of the $p\bar{p}$
distribution amplitudes, time-like versions of GPDs. With sufficient
form factor data at disposal one may attempt a determination of the
time-like GPDs. 
 
\section{Summary}
Results from a first analysis of GPDs at zero skewness have been
discussed. The analysis, performed in analogy to those of the usual
parton distributions, bases on a physically motivated parameterization
of the GPDs with a few free parameters fitted to the available
nucleon form factor data. The analysis of the form factors
provides results on the valence-quark GPDs $H$, $\widetilde{H}$ and
$E$. Interesting results on the structure of the nucleon, in
particular on the distribution of the quarks in the plane transverse
to the direction of the nucleon's momentum are obtained. One finds
that a polarization of the nucleon induces a flavor separation in the
direction orthogonal to the those of the momentum and of the
polarization. An estimate of the average orbital angular momentum the
valence quarks carry lead to a value of $-0.08$. 
 
The zero skewness GPDs are the soft physics input to hard
wide-angle exclusive reactions. For Compton scattering, for instance,
the soft physics is encoded in specific form factors which represent $1/x$
moments of zero-skewness GPDs. Using the results from the GPD analysis
to evaluate these form factors, one can give 
interesting predictions for the differential cross section and 
helicity correlations in Compton scattering. These
predicitons still await their experimental examination.


\begin{thebibliography}{99}
 
\bibitem{mue1994} D.~M\"uller, D.~Robaschik, B.~Geyer, F.~M.~Dittes
and J.~Ho\v{r}ej\v{s}i, 
\Journal{\Fort}{42}{101}{1994} [hep-ph/9812448];
A.~V.~Radyushkin,
\Journal{\PRD}{56}{5524}{1997} [hep-ph/9704207];
      X.~Ji,
\Journal{\PRD}{55}{7114}{1997} [hep-ph/9609381].

\bibitem{CTEQ} J.\ Pumplin {\it et al} [CTEQ collaboration], 
\Journal{\JHEP}{0207}{012}{2002} [hep-ph/0201195].

\bibitem{HERMES} A.~Airapetian {\it et al.}  [HERMES Collaboration],
\Journal{\PRL}{92}{012005}{2004} [hep-ex/0307064].

\bibitem{BB}J.~Bl{\"u}mlein and H.~B{\"o}ttcher,
\Journal{\NPB}{636}{225}{2002} [hep-ph/0203155].

\bibitem{brash} E.J.\ Brash {\it et al},
  \Journal{\PRC}{65}{051001}{2002} [hep-ex/0111038].

\bibitem{DFJK4} M.~Diehl, T.~Feldmann, R.~Jakob and P.~Kroll,
hep-ph/0408173, to be published in {\em Eur. Phys. J.}.

\bibitem{guidal} M.~Guidal, M.~V.~Polyakov, A.~V.~Radyushkin and M.~Vanderhaeghen,
hep-ph/0410251.

\bibitem{SESAM} P.~H\"agler, J.~W.~Negele, D.~B.~Renner, W.~Schroers,
  T.~Lippert and K.~Schilling [LHPC Collaboration],
hep-ph/0410017.

\bibitem{QCDSF} M.~G\"ockeler {\it et al.}  [QCDSF Collaboration],
hep-lat/0410023.

\bibitem{arbarbanel} H.D.I.\ Arbarbanel, M.L.\ Goldberger and S.B.\
  Treiman, \Journal{\PRL}{22}{500}{1969};
P.V.\ Landshoff, J.C.\ Polkinghorne and R.D.\ Short,
\Journal{\NPB}{28}{225}{1971};
M.~Penttinen, M.~V.~Polyakov and K.~Goeke,
\Journal{\PRD}{62}{014024}{2000} [hep-ph/9909489].
\bibitem{barone} V.\ Barone {\it et al.},
 \Journal{\ZPC}{58}{541}{1993};
J.~Bolz and P.~Kroll,
\Journal{\ZPA}{356}{327}{1996} [hep-ph/9603289].

\bibitem{DFJK1} M.~Diehl, T.~Feldmann, R.~Jakob and P.~Kroll,
\Journal{\EPJC}{8}{409}{1999} [hep-ph/9811253].

\bibitem{DFJK3} M.~Diehl, T.~Feldmann, R.~Jakob, and P.~Kroll,
\Journal{\NPB}{596}{33}{2001},
Erratum-ibid.\ {605}{647}{2001}, [hep-ph/0009255].

\bibitem{rad98} A.~V.~Radyushkin,
\Journal{\PRD}{58}{114008}{1998} [hep-ph/9803316].

\bibitem{DY} S.~D.~Drell and T.~M.~Yan,
\Journal{\PRL}{24}{181}{1970}.

\bibitem{ji97} X.-D. Ji,
\Journal{\PRL}{78}{610}{1997} [hep-ph/9603249].

\bibitem{burk} M.~Burkardt,
\Journal{\PRD}{62}{071503}{2000},
Erratum-ibid.\ {66}{119903}{2002}, [hep-ph/0005108],
\Journal{\IJMPA}{18}{173}{2003} [hep-ph/0207047].



\bibitem{hanwen} H.~W.~Huang and P.~Kroll,
\Journal{\EPJC}{17}{423}{2000} [hep-ph/0005318];
H.~W.~Huang, R.~Jakob, P.~Kroll and K.~Passek-Kumericki,
\Journal{\EPJC}{33}{91}{2004} [hep-ph/0309071].

\bibitem{HKM} H.~W.~Huang, P.~Kroll and T.~Morii,
\Journal{\EPJC}{23}{301}{2002}[Erratum-ibid.\ C {\bf 31}, 279 (2003)] 
[hep-ph/0110208].

\bibitem{miller} G.~A.~Miller,
Phys.\ Rev.\ C {\bf 69}, 052201 (2004)
[nucl-th/0402092].

\bibitem{DFHK} M.~Diehl {\it et al.},
\Journal{\PRD}{67}{037502}{2003}
[hep-ph/0212138].

\bibitem{nathan} E99-114 JLab collaboration,
spokespersons C.\ Hyde-Wright, A.\ Nathan and B.\ Wojtsekhowski.

\bibitem{hamilton} D.~J.~Hamilton {\it et al.}  [Jefferson Lab Hall A Collaboration],
nucl-ex/0410001.

\bibitem{DKV} M.~Diehl, P.~Kroll and C.~Vogt,
\Journal{\EPJC}{26}{567}{2003}
[hep-ph/0206288].

\end{thebibliography}
\end{document}